\documentclass[12pt]{iopart}
\usepackage{graphicx}
\usepackage{subfigure}

\begin{document}

\newcommand{\dphi}{$\Delta\phi$ }
\newcommand{\mom}{$p_T$ }
\newcommand{\gevc}{GeV/$c$ }
\newcommand{\flow}{$v_2$ }

\title{Identified Particle Jet Correlations from PHENIX}

\author{Anne Sickles (for the PHENIX Collaboration
\footnote{
For the full list of PHENIX authors and acknowledgments, see
Appendix Collaborations of this volume})}

\address{Brookhaven National Laboratory
Upton, NY 11973}
%\ead{anne@bnl.gov}
\begin{abstract}
Two-particle azimuthal correlations have been shown
to be a powerful probe for extracting novel features of the
interaction between hard scattered partons and the medium
produced in Au+Au collisions at RHIC.  At intermediate $p_T$, 2-5GeV/c,
jets have been shown to be significantly modified in both 
particle composition and  angular distribution compared to p+p
collisions.  
We present
recent PHENIX results from Au+Au collisions for a variety of $p_T$
and particle combinations. 

\end{abstract}

One of the most surprising results from the Relativistic Heavy Ion
Collider (RHIC) has been the large increase in the $p/\pi^+$ and
$\bar{p}/\pi^-$ ratios at intermediate $p_T$, 2--5\gevc~\cite{ppg015}.
Studies of the yields of
  $\Phi$ mesons~\cite{ppg016} and $\Lambda$ baryons~\cite{starlambda}
have indicated that the origin of the excess is related to
the number of valence quarks rather than particle mass.  
Baryon and meson differences have also been studied by measuring
the elliptic flow, $v_2$, of identified particles which have also been shown
to scale with the number of valence quarks~\cite{ppg022,ppg062,starqn}.
In this same \mom range in p+p collisions the dominant particle production
mechanism
shifts from soft, non-perturbative processes to hard parton-parton
scattering followed by jet fragmentation~\cite{starxt}.
The valence quark dependence of these effects has inspired a class
of models based quark recombination; hadronization is modeled
not by fragmentation, but by quarks close together in phase
space coming together to form hadrons
In some of these models intermediate \mom hadrons
primarily come from soft quarks~\cite{friesprc}.  In other
models quarks from jet fragmentation are allowed to recombine 
with soft quarks~\cite{hwa1}. 
All models in this class
extend the  range of soft particle production to higher
\mom and start
with thermalized quark degrees of freedom.

Two-particle correlations can be used to determine whether
the baryon excess is associated with hard or soft processes and
to explore in detail baryon/meson differences at low and intermediate
$p_T$.  A systematic study of these correlations will  
allow discrimination between different hadronization scenarios
and measurement of the role of hard scattering at intermediate $p_T$.
Here we present a selection of the recent PHENIX results
from these correlations.

%Two-particle correlations have been used extensively to
%measure jet properties at intermediate
%\cite{starb2b,ppg033,starlowpt,ppg039,ppg029,ppg072,ppg067} at RHIC.
%\mom at RHIC.
Correlations are measured between two classes
of particles, {\it triggers}
and {\it associated particles}.  
The data presented here are from the 2004 Au+Au $\sqrt{s_{NN}}$=200~GeV RHIC run.
Charged tracks are reconstructed
in the drift chambers and particle identification is done via time of flight.
The start time is from the Beam-Beam Counters and the stop 
time is measured by either the high resolution time of flight or
the lead-scintillator electromagnetic calorimeter, which provide $K/p$
separation to $\approx$4.0~\gevc and $\approx$2.5~GeV/$c$, respectively.
Azimuthal angular differences between 
trigger and associated particles, \dphi, are calculated and  
the non-uniform \dphi acceptance is corrected for with mixed
pairs where the particles are from different events.
Two methods are used to subtract combinatoric pairs from
the underlying event.
The first, the absolute subtraction method, uses
a convolution of the single particle rates to determine the combinatoric
pair rate.  There is an additional correction for the width of the centrality 
bins used ~\cite{ppg033}.  The second method, 
zero yield at minimum (ZYAM),
makes the assumption that there is a region in $\Delta\phi$ where there is
the jet yield is zero~\cite{ppg032}.
Correlations from elliptic flow are 
removed by using \flow values measured independently in PHENIX~\cite{ppg072}.  The
remaining yield
is attributed jet correlations, $J(\Delta\phi)$.  
The \dphi distributions are then described by:
\begin{equation}
\frac{1}{N_{trig}}\frac{dN}{d\Delta\phi} = 
B(1 + 2 v_2^{trig} v_2^{assoc} \cos (2\Delta\phi)) + J(\Delta\phi)
\label{eqflow}
\end{equation}
where $B$ is the combinatoric background level and $v_2^{trig}$ and
$v_2^{assoc}$ are \flow values for triggers and associated particles,
respectively.  $N_{trig}$ is the total number of triggers. 

%\begin{figure}
%\centering
%\includegraphics[width=0.6\textwidth]{ppbar.eps}
%\label{ppbar}
%\end{figure}

\begin{figure}
\begin{minipage}{0.5\textwidth}
\includegraphics[width=\textwidth]{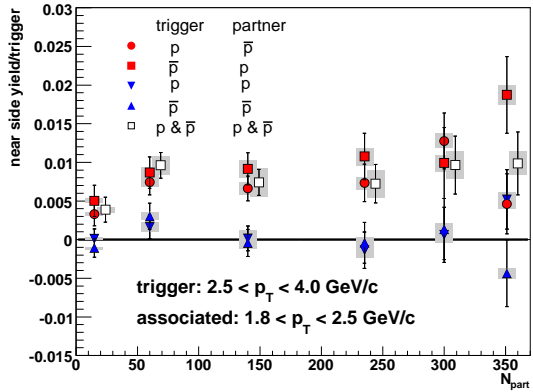}
\end{minipage}
\hspace{\fill}
\begin{minipage}{0.5\textwidth}
\caption{\label{ppbar}Near side conditional yields per trigger for charge selected
 and charge inclusive  $p$ and $\bar{p}$
correlations.  Triggers have 2.5$<p_T<$4.0~\gevc and associated particles
have 1.8$<p_T<$2.5~GeV/$c$.  Error bars are statistical
errors and shaded boxes show systematic errors.  There is an
11.4\% (8.9\%) additional normalization error on baryon ($p$,$\bar{p}$) associated
particle points. Figure is from Ref.~\cite{ppg072}.}
\end{minipage}
\end{figure}

Fig. \ref{ppbar} shows $J(\Delta\phi)$ integrated for $\Delta\phi<0.94$~rad,
i.e the yield of associated particles per trigger as a function of
the number of participating nucleons, $N_{part}$.
The \dphi region covers where two correlated  particles are expected to come from 
the fragmentation of the same jet. Both particles are identified
as $p$ or $\bar{p}$ and different sets of points show
the different charge combinations  with triggers
in the \mom region of the baryon excess.  The baryon-baryon yield
is flat with $N_{part}$,
 except for a smaller yield in the most peripheral collisions.
Same sign pairs, $p$-$p$ and 
$\bar{p}$-$\bar{p}$ (triangles), show no yield 
and opposite sign pairs (filled circles and squares)
are consistent with the charge independent yields.
Yields are comparable for $p$ and $\bar{p}$ triggers.

%\begin{figure}
%\centering
%\includegraphics[width=0.3\textwidth]{assoc_ratios.eps}
%\caption{Ratio of associated baryons to associated mesons as
%a function of \mom for three centrality selections.  Trigger
%particles are charged hadrons with 2.5$<p_T<$4.0~\gevc.  
%Dashed line shows the single particle baryon to meson ratio
%at \mom=1.85\gevc for central collisions calculated from Ref.~\cite{ppg026}.}
%\label{ratio_away}
%\end{figure}

\begin{figure}
\begin{minipage}{0.45\textwidth}
\includegraphics[width=\textwidth]{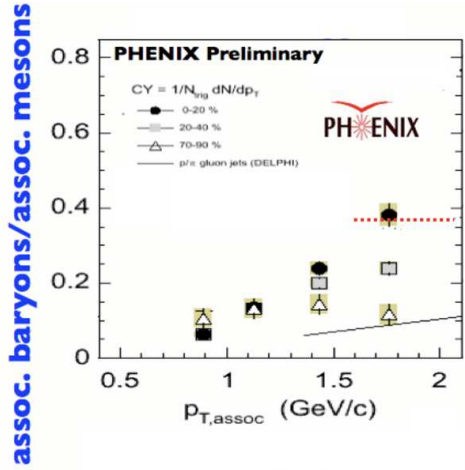}
\end{minipage}
\hspace{\fill}
\begin{minipage}{0.55\textwidth}
\caption{\label{ratio_away} Ratio of associated baryons to associated mesons as
a function of \mom for three centrality selections.  Trigger
particles are charged hadrons with 2.5$<p_T<$4.0~GeV/$c$.  
Dashed line shows the single particle baryon to meson ratio
at \mom=1.85\gevc for central collisions calculated from Ref.~\cite{ppg026}.}
\end{minipage}
\end{figure}

Triggering on an intermediate \mom particle is expected to bias the
near-side jet toward small medium path lengths.  If so, 
the away-side, $\Delta\phi\approx\pi$, typically sees a long medium path length and
could be a sensitive probe of medium modifications to the jet
fragmentation process, hence we measure the particle composition
of the away-side jet.
Fig. \ref{ratio_away} shows the ratio of associated baryons to 
mesons ($\pi^{\pm}$,$K^{\pm}$) with charged hadron triggers,
2.5$<p_T<$4.0~GeV/$c$, as a function of the associated particle $p_T$
integrated over \dphi from $\pi$ to the minimum of $J(\Delta\phi)$.
 In peripheral collisions (triangles) the
 ratio of associated baryons to mesons on the away-side is
approximately flat with $p_T$.
In central collisions (circles) this ratio increases significantly
with the associated particle $p_T$, suggesting that
the away-side jet fragmentation is increasingly baryon rich 
at intermediate $p_T$.  At the highest associated particle
\mom shown in central collisions the ratio of associated baryons to 
mesons is consistent with the value observed in the single particles at
the same \mom and centrality selections~\cite{ppg026}.  

%\begin{figure}
%\centering
%\subfigure[Jet Distribution, 0-20\%]{
%\label{jet_cent}
%\includegraphics[width=0.55\textwidth]{jet_cent.eps}}
%\subfigure[Kurtosis]{
%\label{kurtosis}
%\includegraphics[width=0.40\textwidth]{kurtosis.eps}}
%\caption{(a) $J(\Delta\phi)$ for baryon and meson triggers,  
%1.6$<p_T<$2.0~GeV/$c$, and associated mesons, 1.3$<p_T<$1.6~GeV/$c$.
%Background has been subtracted with the ZYAM assumption. 
%(b) Top panel shows kurtosis as a function of $N_{part}$ in the same \mom range
%as panel (a).  Bottom panel shows the difference in the
%baryon and meson triggered kurtosis.}
%\end{figure}
\begin{figure}[t]
\begin{minipage}{0.55\textwidth}
\includegraphics[width=\textwidth]{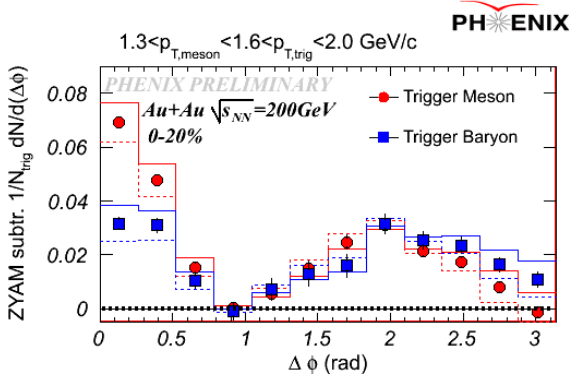}
\end{minipage}
\hspace{\fill}
\begin{minipage}{0.40\textwidth}
\includegraphics[width=\textwidth]{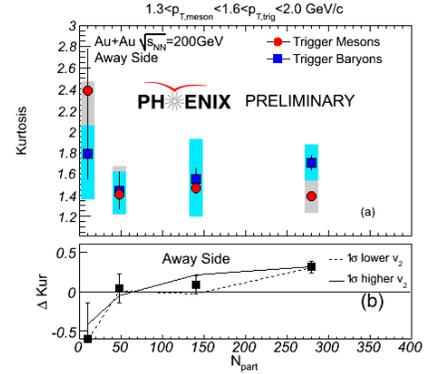}
\end{minipage}
\caption{(a) $J(\Delta\phi)$ for baryon and meson triggers,  
1.6$<p_T<$2.0~GeV/$c$, and associated mesons, 1.3$<p_T<$1.6~GeV/$c$.
Background has been subtracted with the ZYAM assumption. 
(b) Top panel shows kurtosis as a function of $N_{part}$ in the same \mom range
as panel (a).  Bottom panel shows the difference in the
baryon and meson triggered kurtosis.}
\label{kurtosis}
\end{figure}

At intermediate \mom a modified away-side jet shape has also been
observed~\cite{ppg032}.
In contrast to the Gaussian peak centered at $\Delta\phi=\pi$
in p+p collisions, in central
heavy ion collisions the shape is strongly non-Gaussian
and the peak is displaced to $\Delta\phi\approx$2rad.  
Parameters used to characterize the shape in hadron-hadron 
correlations appear to have a universal
$N_{part}$ dependence~\cite{ppg067}.   
Identified particle correlations provide information
on connections between away-side shape modifications
and the baryon excess.
Fig. \ref{kurtosis}(a) shows a \dphi distribution for correlations
of baryon and meson triggers,
1.6$<p_T<$2.0~GeV/$c$, with 
associated mesons, 1.3$<p_T<$1.6~GeV/$c$,
 in central collisions after the \flow
and combinatoric background have been subtracted by the ZYAM
procedure. 
These correlations are at lower \mom
than in Ref.~\cite{ppg067}, but the displaced peak is at
approximately the same position.  
Because of the low trigger \mom and the strong near-side
differences in Fig.~\ref{kurtosis}(a),
further studies are needed to understand baryon/meson differences arising
from the normal fragmentation process.
The top panel of Fig.~\ref{kurtosis}(b)
shows the kurtosis of the away-side peak as a function
of $N_{part}$.  The systematic errors are dominated by
the systematic errors on the \flow values used
and are correlated between baryon and meson triggers.  
$\Delta Kur = Kur_{bar} - Kur_{mes}$ is shown in the bottom 
panel and the lines
show that in central collisions the kurtosis difference
is insensitive to the uncertainties in $v_2$.

We have presented a variety of identified particle correlation
results from PHENIX.  
 Jet shape distributions  at
low \mom in central collisions are different for baryon and
meson triggers.  Mapping out the systematics 
of this effect, particularly the dependence on the associated
particle type, will shed light into the origin
of the modified away-side shape.
Baryons have jet-like correlations in central collisions on both
the near and away-side. 
The weak centrality dependence of the near-side yields in
Fig.~\ref{ppbar} is consistent
with the excess baryons being produced in hard scattering processes.
Other correlations in this \mom range are also consistent with this
picture~\cite{ppg072,ppg033}. 
However, two-particle correlations measure
the average yield per trigger, but not the fraction of
triggers which have associated particles.
 Away side particle ratios in central
collisions are significantly more baryon rich than 
in peripheral collisions.  Recombination models have
not been able to systematically explain the 
two-particle correlation data.  Further theoretical
and experimental studies should allow a better understanding
of the transition from hard to soft physics at intermediate
\mom and the origin of the baryon excess.

%Uncomment for PACS numbers title message
%\pacs{00.00, 20.00, 42.10}
% Keywords required only for MST, PB, PMB, PM, JOA, JOB?
%\vspace{2pc}
%\noindent{\it Keywords}: Article preparation, IOP journals
% Uncomment for Submitted to journal title message
%\submitto{\JPA}
% Comment out if separate title page not required
%\maketitle

\bibliographystyle{iopart-num.bst}
\bibliography{sickles_proceedings}

\providecommand{\newblock}{}
\begin{thebibliography}{10}
\expandafter\ifx\csname url\endcsname\relax
  \def\url#1{{\tt #1}}\fi
\expandafter\ifx\csname urlprefix\endcsname\relax\def\urlprefix{URL }\fi
\providecommand{\eprint}[2][]{\url{#2}}
% Bibliography created with iopart-num v2.0
% /biblio/bibtex/contrib/iopart-num

\bibitem{ppg015}
Adler S~S {\em et~al.\/} (PHENIX) 2003 {\em Phys. Rev. Lett\/} {\bf 91} 172301
  (\textit{Preprint} \eprint{nucl-ex/0305036})

\bibitem{ppg016}
Adler S~S {\em et~al.\/} 2005 {\em Phys. Rev.\/} {\bf 72} 014903
  (\textit{Preprint} \eprint{nucl-ex/0410012})

\bibitem{starlambda}
Adams J {\em et~al.\/} 2004 {\em Phys. Rev. Lett.\/} {\bf 92} 052302

\bibitem{ppg022}
Adler S~S {\em et~al.\/} (PHENIX) 2003 {\em Phys. Rev. Lett.\/} {\bf 91} 182301
  (\textit{Preprint} \eprint{nucl-ex/0305013})

\bibitem{ppg062}
Adare A {\em et~al.\/} (PHENIX) {\em submitted to Phys. Rev. Lett.\/}
  (\textit{Preprint} \eprint{nucl-ex/0608033})

\bibitem{starqn}
Abelev B {\em et~al.\/} (STAR) {\em submitted to Phys. Rev. C\/}
  (\textit{Preprint} \eprint{nucl-ex/0701010})

\bibitem{starxt}
Adams J {\em et~al.\/} 2006 {\em Phys. Lett.\/} {\bf B637} 161--169

\bibitem{friesprc}
Fries R~J {\em et~al.\/} 2003 {\em Phys. Rev.\/} {\bf C68} 044902
  (\textit{Preprint} \eprint{nucl-th/0306027})

\bibitem{hwa1}
Hwa R and Yang C~B 2004 {\em Phys. Rev.\/} {\bf C70} 024904

\bibitem{ppg033}
Adler S~S {\em et~al.\/} (PHENIX) 2005 {\em Phys. Rev\/} {\bf C71} 051902(R)
  (\textit{Preprint} \eprint{nucl-ex/0408007})

\bibitem{ppg032}
Adler S~S {\em et~al.\/} (PHENIX) 2006 {\em Phys. Rev. Lett.\/} {\bf 97} 052301
  (\textit{Preprint} \eprint{nucl-ex/0507004})

\bibitem{ppg072}
Adare A {\em et~al.\/} (PHENIX) {\em submitted to Phys. Lett. B\/}
  (\textit{Preprint} \eprint{nucl-ex/0611016})

\bibitem{ppg026}
Adler S~S {\em et~al.\/} (PHENIX) 2004 {\em Phys. Rev.\/} {\bf C69} 034909
  (\textit{Preprint} \eprint{nucl-ex/0307022})

\bibitem{ppg067}
Adare A {\em et~al.\/} (PHENIX) {\em submitted to Phys. Rev. Lett.\/}
  (\textit{Preprint} \eprint{nucl-ex/0611019})

\end{thebibliography}

\end{document}